\documentclass[twocolumn]{aastex63}

\shorttitle{AASTeX v6.31 Sample article}
\shortauthors{Zapata et al.}
\graphicspath{{./}{figures/}}
\newcommand{\dechms}[4]{$#1^{\rm h}#2^{\rm m}#3\mbox{$^{\rm s}\mskip-7.6mu.\,$}#4$}

\newcommand{\decdms}[4]{$-#1^{\circ}#2'#3\mbox{$''\mskip-7.6mu.\,$}#4$}
\usepackage{graphicx}

\begin{document}

\title{Catching the Butterfly and the Homunculus of $\eta$ Carinae with ALMA}

\correspondingauthor{Luis A. Zapata}

\email{l.zapata@irya.unam.mx}

\author{Luis A. Zapata}
\author{Laurent Loinard}
\affil{Instituto de Radioastronom\'\i a y Astrof\'\i sica, Universidad Nacional Aut\'onoma de M\'exico, P.O. Box 3-72, 58090, Morelia, Michoac\'an, M\'exico}

\author{Manuel Fern\'andez-L\'opez}
\affiliation{Instituto Argentino de Radioastronom\'\i a (CCT-La Plata, CONICET; CICPBA), C.C. No. 5, 1894, Villa Elisa, Buenos Aires, Argentina}

\author{Jesús A. Toalá}
\author{Ricardo F. González}
\affil{Instituto de Radioastronom\'\i a y Astrof\'\i sica, Universidad Nacional Aut\'onoma de M\'exico, P.O. Box 3-72, 58090, Morelia, Michoac\'an, M\'exico}

\author{Luis F. Rodr\'\i guez}
\affil{Instituto de Radioastronom\'\i a y Astrof\'\i sica, Universidad Nacional Aut\'onoma de M\'exico, P.O. Box 3-72, 58090, Morelia, Michoac\'an, M\'exico}
\affil{Mesoamerican Centre for Theoretical Physics, Universidad Aut\'onoma de Chiapas, Carretera Emiliano Zapata Km. 4 Real del Bosque, 29050 Tuxtla Guti\'errez, Chiapas, M\'exico}

\author{Theodore R. Gull}
\affil{NASA Goddard Space Flight Center, Exoplanets and Stellar Astrophysics Laboratory, Code 667, Greenbelt, MD 20771} 

\author{Patrick W. Morris} 
\affil{California Institute of Technology, IPAC, M/C 100-22, Pasadena, CA 91125, USA}

\author{Karl M. Menten}
\affil{Max-Planck-Institut f\"ur Radioastronomie, Auf dem H\"ugel 69,  D-53121 Bonn, Germany} 

\author{Tomasz Kamiński}
\affil{Nicolaus Copernicus Astronomical Center of the Polish Academy of Sciences, Toruń, Poland}



\begin{abstract}
The nature and origin of the molecular gas component located in the circumstellar vicinity of $\eta$ Carinae are still far from being completely understood. 
Here, we present Atacama Large Millimeter/Submillimeter Array (ALMA) CO(3$-$2) observations with a high angular resolution ($\sim$0.15$''$), and 
a great sensitivity that are employed to reveal the origin of this  component in $\eta$ Carinae. These observations reveal much 
higher velocity ($-$300 to $+$270 km s$^{-1}$) blue and redshifted molecular thermal emission than previously reported, which we associate with the lobes of the Homunculus Nebula, 
and that delineates very well the innermost contours of the red- and blue-shifted lobes likely due by limb brightening. The inner contour of the redshifted emission
was proposed to be a {\it disrupted torus}, but here we revealed that it is at least part of the molecular emission originated from the lobes and/or the expanding equatorial skirt.
On the other hand, closer to systemic velocities ($\pm$100 km s$^{-1}$), the CO molecular gas traces an inner butterfly-shaped structure 
that is also revealed at NIR and MIR wavelengths as the region in which the shielded dust resides. The location and kinematics of the molecular component indicate that this material 
 has formed after the different eruptions of $\eta$ Carinae.
\end{abstract}


\section{introduction} \label{sec:intro}

$\eta$ Carinae is one of the most studied evolved and massive stars in the entire sky.  Classified as a Luminous Blue Variable (LBV), $\eta$ Carinae
is an eccentric ({\it e} = 0.9) massive binary system composed of a hypergiant blue star ($\eta$ Carinae A) of 90 M$_\odot$ and a hotter, less massive 
companion ($\eta$ Carinae B, which is likely in a Wolf-Rayet phase) of about 30$-$40 M$_\odot$ \citep{dam2019,gro2012,pit2002,teo2016}. 
During the 1840s, $\eta$ Carinae experienced a dramatic outburst that is nicknamed {\it The Great Eruption} which ejected  {\it{at least}} 40 M$_\odot$ into 
the ISM \citep{mor2017}, forming the well-known Homunculus Nebula \citep{gav1950}. 

The Homunculus Nebula is characterized by its expanding bipolar southeast-northwest morphology \citep[with radial velocities 
reaching $\sim$ 650 km s$^{-1}$ at the pole zones, and with an angular projected size of about 20$''$ $\times$ 10$''$, see Figure 1 of][]{gon2010}, 
and a high content of  warm dust, which absorbs much of the optical light that is re-emitted at NIR (near-IR) wavelengths. 
The Homunculus nebula has been proposed to be the result of an extremely enhanced stellar wind prior to the merger of a massive close binary 
that produced the 90 M$_\odot$  component \citep{por2016,nat2018,hir2021}. This merger was stimulated by the gravitational interaction of what 
was then the third star in the system and now is the 30 M$_\odot$ companion \citep{por2016}.


\begin{figure*}[ht!]
\epsscale{1}
\plotone{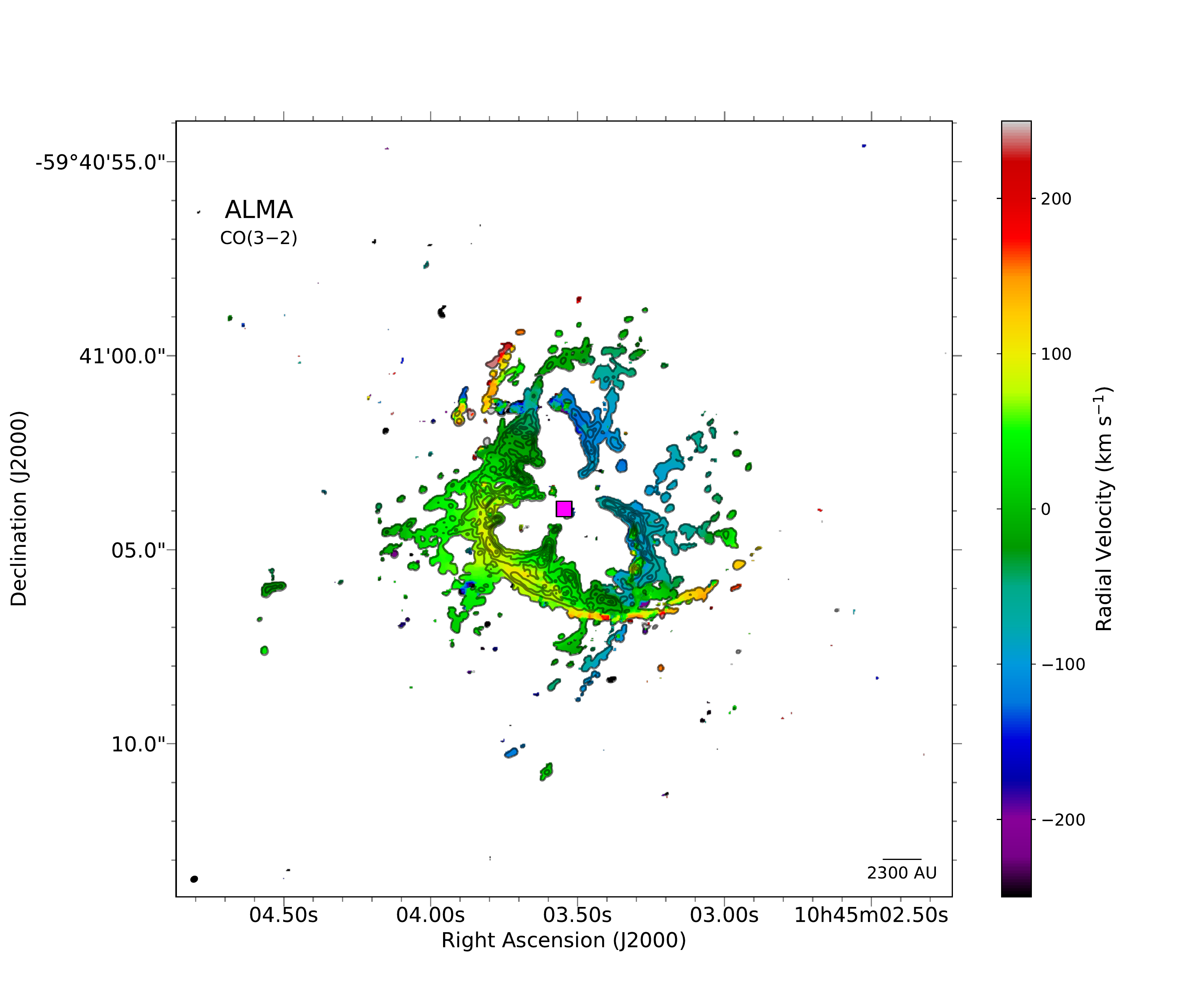}
\caption{\scriptsize ALMA CO(3$-$2) moment zero (contours) and one (color scale) maps of the $\eta$ Carinae region.  
The moment zero is the integrated value of the spectrum, while the moment one map is the intensity weighted coordinate. 
To compute this moment zero map, we integrated the observed flux density over radial velocities from $-$300 to $+$270 km s$^{-1}$. 
The contours are from  $-$10\%, 10\%, 30\%, 50\%, 70\% and 90\% of the peak emission. The peak of the millimeter CO line emission is 0.2 Jy Beam$^{-1}$ km s$^{-1}$.    
The half-power contour of the synthesized beam of the line image is shown in the bottom-left corner. The LSR radial velocity scale-bar (in km s$^{-1}$) 
is shown at the right. In the bottom right-corner, the spatial scale is indicated for a distance of 2.3 kpc \citep{smi2008}.   
The magenta square marks the location of the binary optical star $\eta$ Carinae \citep{hog2000}. The pixel size is 0.03 arcsec.
\label{fig:fig1}}
\end{figure*}

 At (sub)millimeter wavelengths, several molecules have been reported to be associated with the circumstellar vicinities of  $\eta$ 
 Carinae \citep[{e.g.,}][]{gull2020}.  \citet{loi2012} reported the thermally excited CO, CN, HCO${^+}$, HCN, HNC, and N$_2$H$^+$, 
 as well as several of their $^{13}$C isotopologues observed with the  Atacama Pathfinder Experiment (APEX) 12 m single dish submillimeter 
 telescope.  In that study, the relative intensities of the multiple CO lines suggested that the molecular emission 
 was more compact than the APEX beam ($9''$--$18''$ Full Width Half Maximum {\it i.e.} FWHM), and probably was concentrated very close to the Homunculus Nebula. 
 Subsequent interferometric observations of the HCN (1$-$0) line with the Australia Telescope Compact Array (ATCA) confirmed that the thermal molecular 
 emission is indeed concentrated in the central few arcseconds around $\eta$ Carinae, and revealed a clear east-west velocity gradient of 150 km s$^{-1}$\citep{loi2016}. 
 Furthermore, more recent ALMA observations are interpreted to trace a disrupted molecular torus around $\eta$ Carinae, with expanding radial velocities 
 of approximately $\pm$200 km s$^{-1}$, and which seems to be aligned well with the equatorial warm dust around  $\eta$ Carinae previously 
 observed in the MIR (mid-IR) and NIR images \citep{nat2018}. Similar molecular structures were later reported by \citet{Mor2020} 
 using again ALMA observations, who advocated the view that  the properties of the dust and CO emission may be measurably sensitive to 
 changes in the amount of ultraviolet radiation escaping from the binary during periastron passage.
 In these ALMA CO(2$-$1) images, the disrupted molecular torus did not show its typical 
 northwest aperture as reported in \citet{nat2018}. This was attributed to the sensitivity differences between the synthesized ALMA observations \citep{Mor2020}. 
 \citet{bor2019} reported an asymmetric and more compact ($\leq$ 1$''$) molecular 
 structure (traced by the HCO$^+$) to the northwest of $\eta$ Carinae, and suggested that it likely was expelled in a second eruption. 
 However, \citet{abr2020} proposed that this compact emission was instead arising from ionized gas traced by the recombination 
 line H$40\delta$, not from the molecular line HCO$^+$. 
 
 In the 1890s, $\eta$ Car underwent a second eruptive event ({\it the minor eruption}) from which an internal nebula was produced \citep[see, for instance,][]{ish2003,smi2004}. 
 The so-called {\it little Homunculus} extends $\pm$ 2$''$ along the symmetry axis, and expands currently at a speed of $\sim$ 250 km s$^{-1}$. 
 It is worth mentioning that the polar caps of the {\it little Homunculus} are not similar in any way to the external lobes of the {\it large Homunculus}.  
 The polar caps in both structures are very different in position and excitation. 
 The total mass expelled from the primary star during the minor event of about 0.1 M$_{\odot}$, and a total kinetic energy of 10$^{46.5}$ erg was released, 
 which is a factor of 10$^{3}$ smaller than the corresponding value of the major eruption (10$^{49.7}$ erg).
 
 In this study, we present ALMA CO(3$-$2) observations with high angular resolution, and a superb sensitivity that reveal the innermost morphology, 
 and kinematics of the multiple molecular structures associated with  $\eta$ Carinae, and its real angular size. 
 
 \begin{figure}[ht!]
\epsscale{1.15}
\plotone{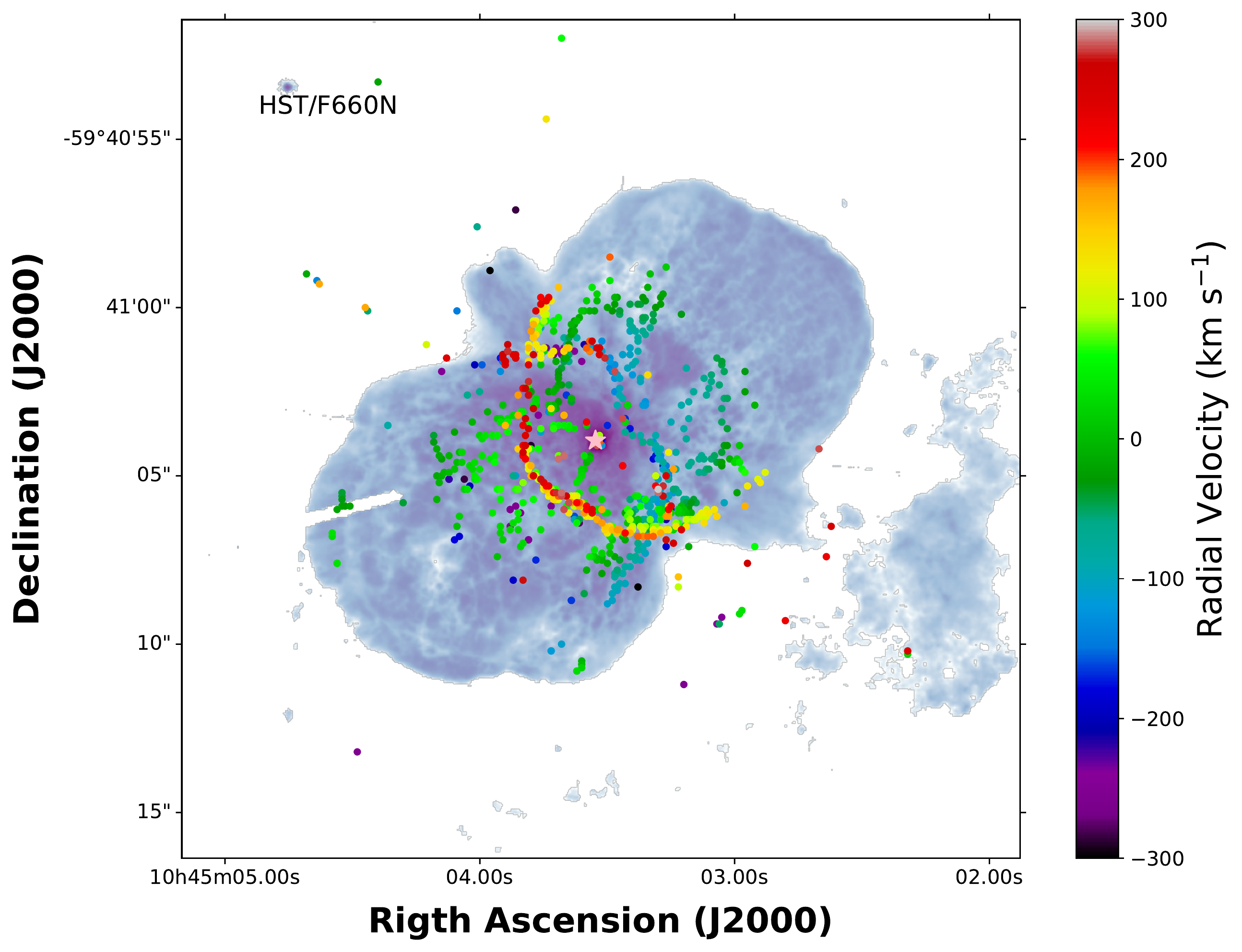}\\
\plotone{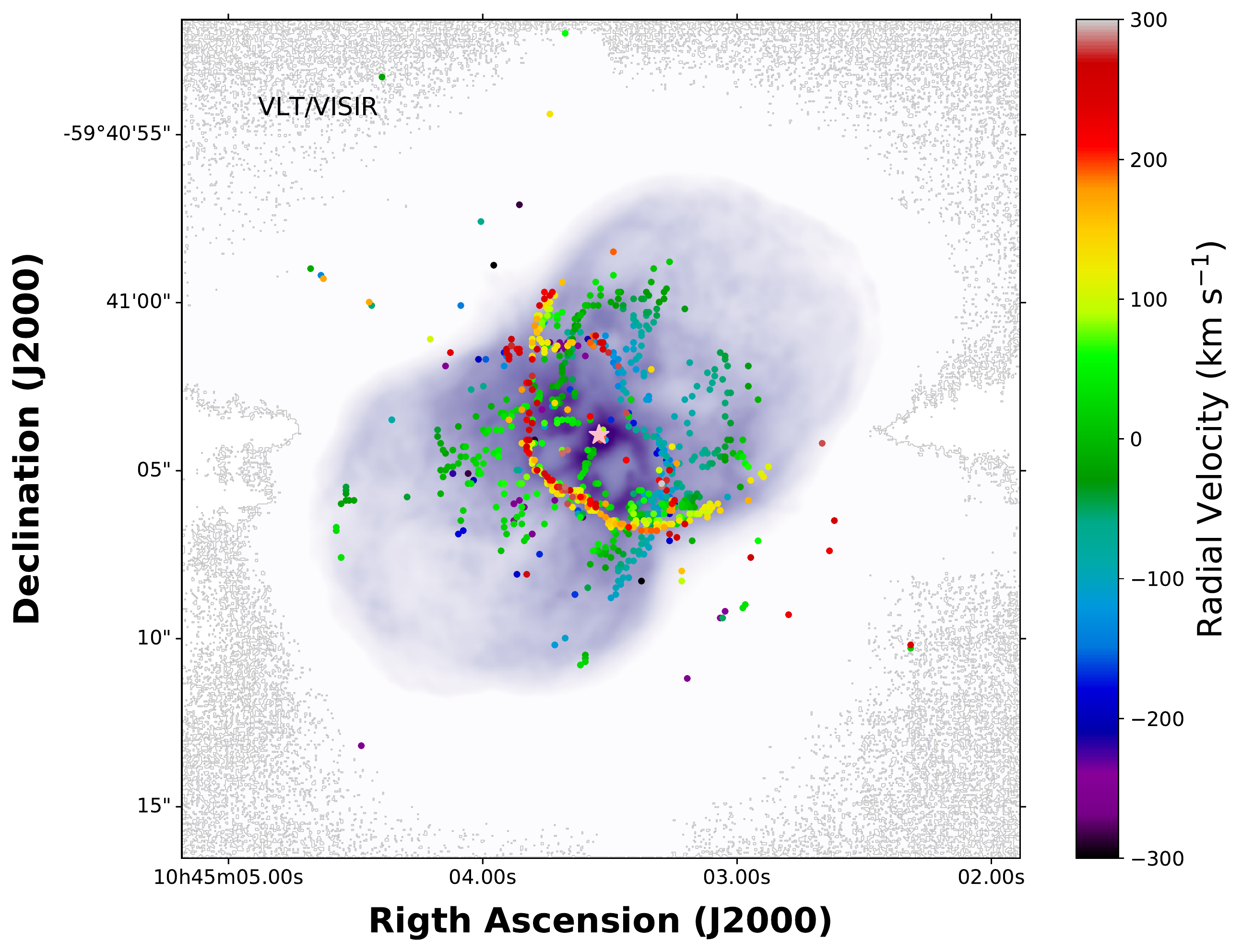}
\caption{\scriptsize  {\it Upper Panel:} Archive optical HST/ACS (F660N) image (color scale) of the $\eta$ Carinae Homunculus overlaid with the positions 
of every CO(3$-$2) condensation in the channel velocity cube. 
To compute this map, we used the molecular emission located in radial velocities from $-$300 to $+$270 km s$^{-1}$ for the blueshifted  and redshifted positions. 
The LSR radial velocity scale-bar (in km s$^{-1}$) is shown at the right. The pink star marks the location of the binary optical star $\eta$ Carinae \citep{hog2000}.
We shifted in position the optical image to be aligned with the ALMA CO image.   {\it Lower Panel:}  MIR 7.9 $\mu$m VLT/VISIR \citep{men2019} image 
(color scale) of the $\eta$ Carinae Homunculus overlaid with the positions of every CO(3$-$2) condensation in the channel velocity cube. Symbols are as in the upper panel.
\label{fig:fig2}}
\end{figure}

\begin{figure*}[ht!]
\epsscale{1.15}
\plotone{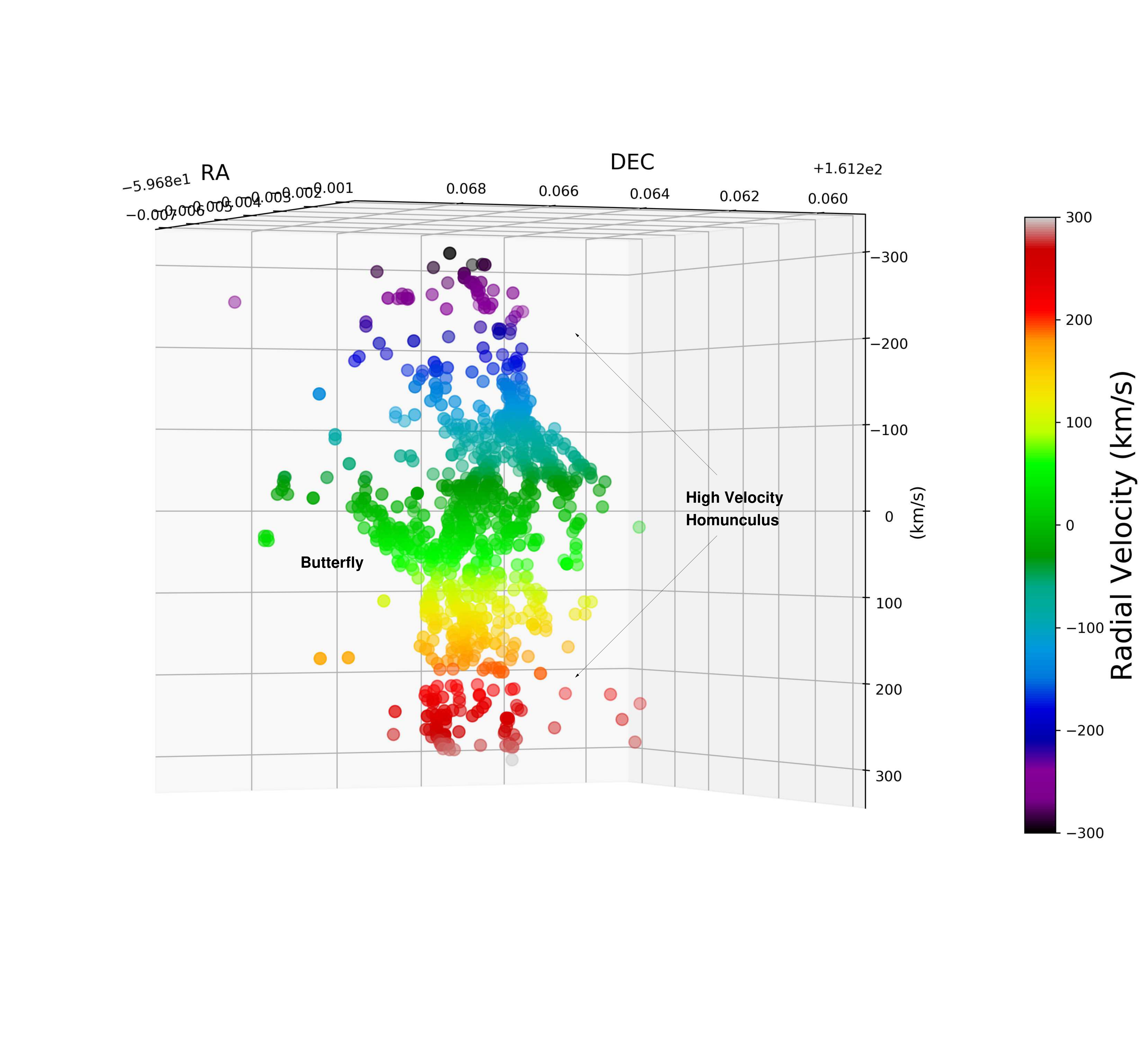}
\vspace{-2cm}
\caption{\scriptsize Three-dimensional illustration of the ALMA CO(3$-$2) molecular emission from $\eta$ Carinae. The blue and red colors represent the 
blueshifted and redshifted emission, respectively.   The LSR radial velocity scale-bar (in km s$^{-1}$) 
is shown at the right. In the animation the magenta square marks the location of the binary optical star $\eta$ Carinae \citep{hog2000}. 
The third dimension (radial velocity) corresponding to the expansion of the Homunculus since outburst in the 1840s.
There is a 3D animation associated to this illustration. In this animation, the third dimension is the radial velocity in km $^{-1}$ and the other two are position (RA and Dec).  
Here, we are assuming a Hubble-Law expansion. 
\label{fig:fig3}}
\end{figure*}

\section{observations} \label{sec:obs}

The observations were carried out with 41 antennas of ALMA 
in 2016 October during the Cycle 4 science data programme 2016.1.00585.S (P.I. Gerardo Pech-Castillo). These observations only included
12 m diameter antennas, which yielded projected baselines from 18.6 to 1800 m (14.3 $-$ 1384 k$\lambda$).  The FWHM of the primary beam is 
16.5$''$ at this frequency (Band 7), so that the bulk of the molecular and dusty material associated with $\eta$ Carinae is well-covered \citep[{\it e.g.} ][]{loi2016,men2019}.  
We are aware that maybe some faint molecular structures far from the central region of $\eta$ Carinae could be lost; however, a comparison 
between the APEX \citep{loi2016}, and ALMA CO(3$-$2) spectra does not reveal an extended component for this source, see the Appendix, Figure \ref{fig:fig6}.     

The phase center was located in the sky position $\alpha_{J2000.0}$ = \dechms{10}{45}{03}{522}, and $\delta_{J2000.0}$ = \decdms{59}{41}{03}{88} with a 
 total integration time on source of 11 min. The maximum recoverable scale for these observations is 1.5$''$ although the minimum baseline corresponds 
 to an angular scale of  $\sim$14$''$.

With an average system temperature of about 140 K and with precipitable water vapor column around 0.57 mm, the observations were made in optimal
conditions. In order to reduce the atmospheric phase fluctuations during the observations, the 183 GHz water line was monitored with water 
vapor radiometers (WVR). The quasars J0538$-$4405, J1107$-$4449, J1047$-$6217, and J1032$-$5917 were used as the amplitude, atmosphere, bandpass, 
pointing, gain fluctuations, and WVR calibrators. Some of the quasar scans were repeated for different calibrations.   

The ALMA digital correlator was configured in four spectral windows of width 937.5 MHz and 3840/1920 channels each. 
This provides a channel spacing of  244.141 kHz ($\sim$0.21 km s$^{-1}$) and 488.281 kHz ($\sim$0.42 km s$^{-1}$), respectively. 
However, given the broad range of the CO velocities ($-$ 300 to $+$ 270 km s$^{-1}$), we smoothed the channel spacing to 5 km s$^{-1}$. 
The four spectral windows were centered at rest frequencies of  345.817 GHz, 345.362 GHz, 354.859 GHz, and 356.757 GHz
intended to detect four rotational lines, CO (3$-$2), H$^{13}$CN (4$-$3), HCN (4$-$3), and HCO$^{+}$ (4$-$3).  All these
spectral molecular lines have been already reported to be present in the vicinities of $\eta$ Carina \citep{loi2012,loi2016,bor2019}.   
In this paper, we concentrated only on the CO(3$-$2) emission, the remaining molecules have already been studied by \citet{bor2019} or will be
presented in a more dedicated future study.  We have revised very carefully  the spectra of the CO(3$-$2) to avoid a possible contamination of the 
H$^{13}$CN (4$-$3), especially at high positive velocities ($\geq$ 270 km s$^{-1}$). However, we do not find such contamination.  But, there could be some faint H$^{13}$CN (4$-$3)
line contamination in the spectral channels in the range between 210 and 270 km s$^{-1}$, and in a very low level.  

We used the  Common Astronomy Software Applications (CASA) package, Version 4.7, to calibrate, image, and analyze the ALMA data. 
We also used some routines in Python to image the data \citep{ast2013}. The task TCLEAN in CASA was used for imaging the data. 
We set the {\tt\string Robust} parameter of TCLEAN equal to 0.5.  

The line image {\it rms}-noise is 2.5 mJy beam$^{-1}$ km s$^{-1}$ at an angular resolution of 0.17$''$ $\times$ 0.13$''$ with a PA of $-$59$^\circ$. 
The ALMA theoretical rms noise for this configuration, integration time, bandwidth (channel spacing), and frequency is about 2.2 mJy 
beam$^{-1}$ km s$^{-1}$, which is also very close to the value we obtain in the line images. Phase self-calibration was done using the continuum emission 
as a model, and then we applied the solutions to the line emission. We obtained about a factor of ten of improvement in the continuum {\it rms}-noise, 
and a factor of seven of improvement in the channel line {\it rms}-noise. These noises levels allow us to reveal new CO features associated with 
$\eta$ Carinae as compared to the ALMA CO (3$-$2) maps presented in \citet{bor2019} with a channel cube {\it rms}-noise of 15 mJy beam$^{-1}$ km s$^{-1}$
or in \citet{Mor2020} with a channel CO(2$-$1) cube {\it rms}-noise of 32 mJy beam$^{-1}$ km s$^{-1}$.

\begin{figure*}[ht!]
\epsscale{1.0}
\plotone{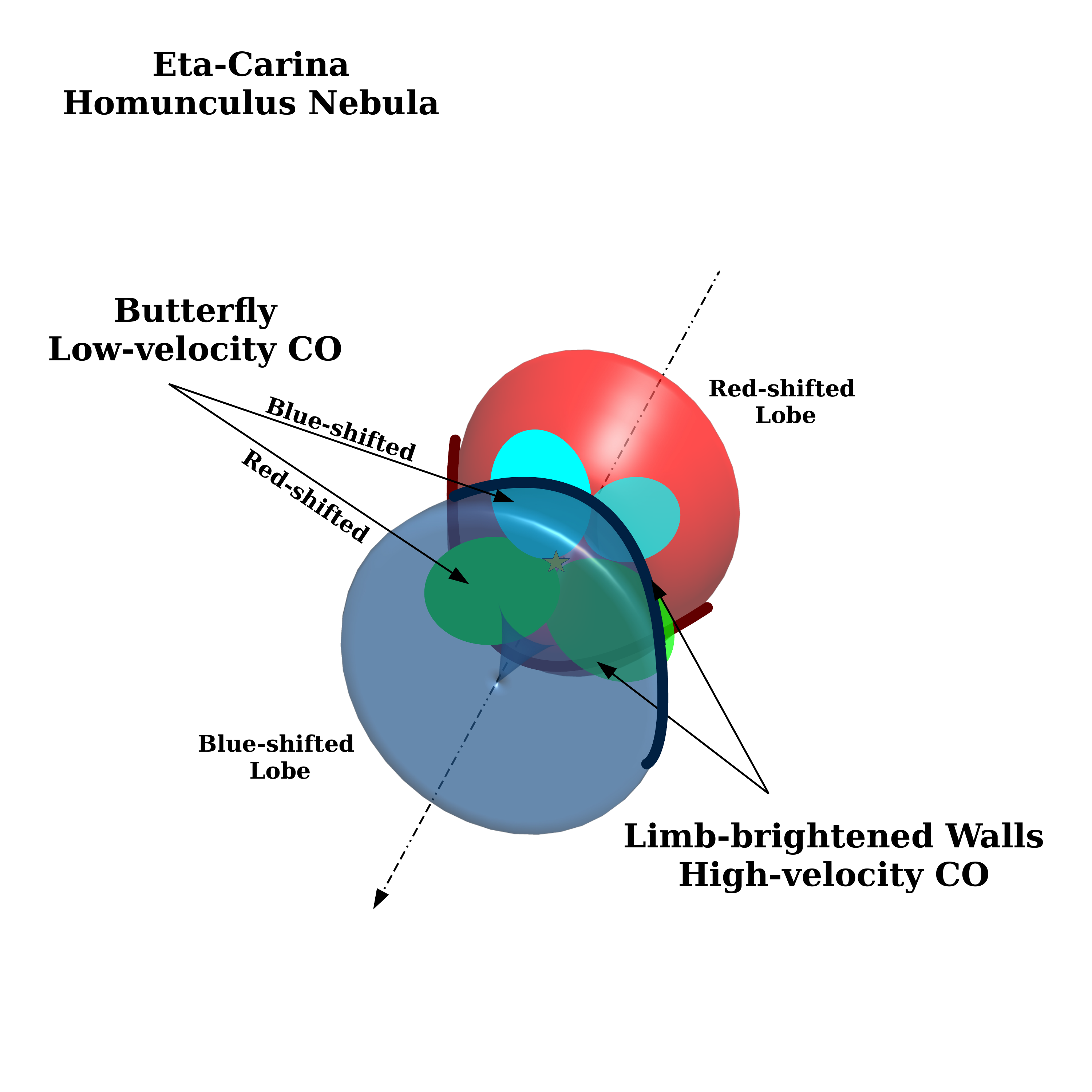}
\caption{\scriptsize  Sketch of the Homunculus and the Butterfly structures in the $\eta$ Carinae system.  
\label{fig:fig4}}
\end{figure*}

\section{Results and Discussion} \label{sec:dis}

In Figure \ref{fig:fig1},  we present the moment zero and one maps of the CO(3$-$2) emission detected with ALMA in the surroundings of $\eta$ Carinae.
This map was computed using only radial velocities from $-$300 to $+$270 km s$^{-1}$. Some compact molecular features are seen beyond these velocities, and
are localized far from $\eta$ Carinae,  possibly originating from faint molecular lines far from the CO rest frequency. 
In particular, some of these compact features with not so high-velocity  ($-$300 km s$^{-1}$ $\leq$ Vel. $\leq$ $+$270 km s$^{-1}$) are mostly associated 
with the Homunculus Nebula as we will demonstrate with the next set of images. These high-velocity structures are not seen in the recent ALMA maps, 
as for example in \citet{nat2018}, who detected strong CO emission only in a reduced velocity window from $-$100 to $+$150 km s$^{-1}$. 
Their channel cube {\it rms}-noise is about a factor 10 higher (21.6 mJy beam$^{-1}$ km s$^{-1}$).
Moreover, \citet{mor2017} also detected hotter CO velocity emission with Herschel/HIFI only over the window of $-$150 to $+$150 km s$^{-1}$ 
at main beam rms noise of $\approx$0.06 K. Our maps reveal a mix of radial velocities with blue- 
and red-shifted CO emission in both sides of the nebula, and with an elongated structure in the direction of the Homunculus Nebula and with an angular 
size of about 15$''$ $\times$ 10$''$. This is a bit different from what is reported in \citet{nat2018} or \citet{Mor2020}, who find lower-velocity redshifted components to be 
mostly localized at the southeast side, and the blueshifted velocities are in the northwest side of the molecular disrupted roundish torus or shell.  
This is likely due to our better sensitivity, and high angular resolution, which allow us to trace and resolve fainter structures. 
Three main features are revealed from this figure, we describe them as follows:

\begin{itemize}

\item A thin redshifted well-defined open arc-like structure in the southeast of $\eta$ Carinae, and that extends to its northwest. It is open toward the NW. 
This structure has the largest redshifted velocities found in this study ($\sim$270 km s$^{-1}$). The thin width of the 
open arc-like structure is about one synthesized beam, that in spatial scales corresponds to about 350 au. This structure has a very good 
correspondence with the redshifted expanding horseshoe or disrupted torus reported in \citet{nat2018}. 
On the other hand, given the extent of this thin arc, the redshifted structure could be tracing the innermost NE lobe of the Homunculus
by limb brightening.  Another possibility for enhanced emission along this geometry is the intersection 
of the skirt with the Homunculus so that higher emission is observed along our line of sight. In Figure \ref{fig:fig7}, in the Appendix, we illustrate the three different structures
possibly associated with the CO emission   

\item The Butterfly structure with four wings well delineated. While the northern wings are blueshifted with respect to the velocities close 
to eta-Carina's systemic LSR velocity, $V_\mathrm{LSR}$ = $-$19.7 km s$^{-1}$ \citep{art2011}, the southern wings are preferentially redshifted.  
Each of these CO wings have more complex internal velocity variations, possibly indicating an intricate 3D structure.
The radial velocities associated with this structure are closer to systemic in a range of $-$100  to $+$100 km s$^{-1}$. 
The edges of the butterfly wings point toward the position of the central star of $\eta$ Carinae.
 The Butterfly is present in all previous ALMA maps \citet{nat2018,Mor2020,bor2019}.

\item Faint filaments with velocities that are blueshifted relative to and close to $\eta$ Carinae's systemic velocity are located in 
the southeast side of the star. The filaments extend over roughly a few arseconds, and are far from the Butterfly nebula. The filaments are 
present at both southern edges of the nebula.

\end{itemize}

In Figure \ref{fig:fig2}, we have overlaid the CO emission on the optical and infrared images. The optical image was retrieved from the archive 
of the {\it Hubble Space Telescope} using the Advanced Camera for Surveys (ACS) instrument with filter 660N. For the infrared image, we used  
one obtained at 7.9 $\mu$m with the Very Large Telescope using the Imager and Spectrometer for mid-InfraRed (VISIR), this infrared image 
was obtained from \citet{men2019}. The CO data (in form of colored dots) were extracted from the channel spectral cube using a Gaussian fitting 
to every compact condensation employing the MIRIAD task IMSAD, then plotted in both figures. We regarded a compact CO condensation 
as a real entity when its flux density was higher than 8$\sigma$ with $\sigma$ equal to 2.5 mJy beam$^{-1}$ km s$^{-1}$. 
The resulting CO structure agrees very well with that revealed in the moment zero and one (Figure \ref{fig:fig1}).
The low intensity sources have sizes around one synthesized beam.
The three molecular structures described in Figure \ref{fig:fig1}, the disrupted torus, the Butterfly, and the filaments are all clearly discernible in these figures. 
For the case of the disrupted torus, this configuration is clearly tracing 
the edge of the northwest lobe of the Homunculus Nebula.  At optical wavelengths, the northwest lobe shows similar redshifted emission but 
with larger radial velocities that reaches $\sim$ 700 km s$^{-1}$ at the poles \citep{nat2018,art2011}. These structures are tracing the edges of the lobes 
and are very likely produced due to limb-brightening effects.  On the other hand, the blueshifted lobe does not show a similarly coherent structure.
It rather has some partially detected parts of the blueshifted filaments to the southwest of the Homunculus nebula. 
In the channel map of Figure 5 (e.g., at $-$120 km s$^{-1}$) the southwest part of the filament is clearly identified.
These limb-brightened edges likely trace regions in 
which the line-of-sight runs tangentially along a large column of material lying on the walls of the Homunculus Nebula, which, as is well known, 
presents an overall hour-glass morphology. The blueshifted lobe shows a similar structure in the southwest part of the Homunculus 
(see channels with velocities $-$90 to $-$180 km s$^{-1}$ in Figure \ref{fig:fig5}). However, the northeast limb-brightened edge is not so clear in 
CO (e.g., velocity channels from 0 to $-$60 km s$^{-1}$). Instead, there are some shreds of emission or CO etchings in the northeast 
surface of the Nebula, which may be tracing the limb-brightened structures at velocities closer to the LSR.
 
From the optical image it is clear that the CO emission associated with the Butterfly is located in the dark areas of the expanding equatorial structure. 
This is confirmed in the lower panel of  Figure \ref{fig:fig2}, where we show the MIR emission from $\eta$ Carinae. In this panel, one sees an excellent 
correlation between the CO emission in the wings of the Butterfly and the MIR emission. However, the MIR emission is located in the inner 
parts of the wing of the Butterfly, likely due to dust shielding. 

In Figure \ref{fig:fig3}, a three-dimensional view of the CO emission in $\eta$ Carinae is presented. Here, we recall that we assume that the radial 
velocity is proportional to the distance, so the third dimension is considered a space-dimension. This illustration has three axis, 
the Right Ascension (R.A.), the Declination (DEC.), and the radial velocity (in km s$^{-1}$) of the thermal emission. This illustration reveals the 
gas kinematics of the three molecular structures traced by these observations: The Butterfly, The Homunculus Nebula, and the filaments. 
This illustration shows that some regions of the bipolar structure of the Homunculus Nebula are associated to the CO emission. 
The Butterfly is a chain of filaments aligned  with the central position of $\eta$ Carinae. The kinematics found in the wings of the Butterfly 
(lower absolute radial velocities than the CO belonging with the Homunculus) suggest that these structures may lie closer to the equatorial plane 
of the Homunculus, that is, inclined 41$^\circ$ \citep{smi2006} with respect to the plane of the sky. In this scenario, the blueshifted wings seem to 
lie on the northwest part of the equatorial plane (channels from $-$60 to $-$150 km s$^{-1}$), while the redshifted wings would be on the opposite 
southeast side of that same plane (channels from 0 to $+$150 km s$^{-1}$). To illustrate this situation, we show in the Appendix slices of the channel 
velocity cube (see, Figure \ref{fig:fig5}).
   
A cartoon of the different parts and orientations of the Carinae Nebula is presented in Figure \ref{fig:fig4}. In this image, we draw the different molecular 
structures traced by this work and revealed by the ALMA observations.   

In Figure \ref{fig:fig8}, we have computed two Position-Velocity diagrams across and along the Homunculus Nebula and passing over the position of Eta Carinae. These Figures
reveal bipolar structures associated with the molecular emission arising from the Homunculus Nebula.  These structures are also observed in the 3D animation. Similar structures are 
also observed in H$_2$ tracing the Homunculus \citep{smi2004}. 

Finally, assuming that the CO(3$-$2) line emission is optically thin and in local thermodynamic equilibrium, we estimate the outflow mass using the 
following equation \citep[see,][]{geo2020}, adapted for the J=3$-$2 transition:

\scriptsize
$$
\left[\frac{M_{H_2}}{M_\odot}\right]=3.4 \times10^{-16}\,T_\mathrm{ex}\,e^{\frac{33.18}{T_\mathrm{ex}}}
\,X_\frac{H_2}{CO} 
\left[\frac{\int \mathrm{I_\nu dv}}{\mathrm{Jy\,km\,s}^{-1}}\right]
\left[\frac{\theta_\mathrm{maj}\,\theta_\mathrm{min}}{\mathrm{arcsec}^2}\right]
\left[\frac{D}{\mathrm{pc}}\right]^2, 
$$ 
\normalsize

\noindent where we used 2.8 as mean molecular weight, and the abundance ratio between the molecular hydrogen and the carbon monoxide 
\citep[10$^4$, e.g.][]{sco1986}, $T_\mathrm{ex}$ is in units of K and assumed to be 200~K (a value close the dust temperature in the Carinae Nebula), 
$\int \mathrm{I_\nu dv}$ is the average intensity integrated over velocity with a value of 0.1 $\mathrm{Jy\,km\,s}^{-1}$ for the contour of redshifted side of the Homunculus, 
$\theta_\mathrm{maj}$ and $\theta_\mathrm{min}$ are the projected major and minor axes of the outflow lobe, and $D$ is the distance to the source, 2.3 kpc \citep{smi2008}. 
We estimate a mass for the redshifted contour of the Homunculus $\sim$ 0.001 M$_\odot$. This mass value is highly uncertain,
as is discussed in \citet{nat2018}, mainly due to the standard CO/H$_2$ number ratio of 10$^{-4}$ used here. 
Moreover, H$_2$ mass in the nebula seems to be a changing quantity over time, probably related to changes in the UV field.
For example, \citet{not2002} estimated a value for this number ratio (CO/H$_2$) of 1.6 $\times$ 10$^{-4}$ in the circumstellar environment 
of the LBV star AG Carinae. Therefore, we would like to stress the uncertainty 
of this estimate of the molecular gas mass in $\eta$ Carina. In any case, this mass reported here is however negligible compared 
the total mass of the Homunculus Nebula, see  \citet{mor2017}.

Finally, we do not find evidence for a molecular counterpart of the  {\it little Homunculus} reported in \citet{ish2003}.

\section{Summary}

In conclusion, the sensitive and high spatial resolution of the ALMA observations of $\eta$ Carinae have revealed, for the first time, 
higher velocity CO emission than previously reported.  We interpret the arc of emission exhibiting velocities over the range of $+$300 to $-$300 
km s$^{-1}$ to be associated with the bipolar lobes observed in the optical. The CO emission is found to delineate very well the innermost
contour of the redshifted lobe likely due by limb brightening or possibly arising in the intersection of the skirt with the Homunculus in our line 
of sight so that we see higher emission measure.
At systemic velocities the thermal emission from the CO is found close to $\eta$ Carinae and in the expanding equatorial structure 
revealed also at optical wavelengths. The wings of the Butterfly are correlated with the dark dusty patches observed at optical wavelengths and
that are bright in the MIR maps. There is a shift between the wings of the Butterfly mapped in the MIR and CO, likely because the molecules are shielded from 
UV radiation. Additionally, filamentary blueshifted CO emission is also found associated with the  Homunculus Nebula. A map of the 3D molecular 
emission in the vicinities of $\eta$ Carinae is presented and reveals a structured morphology associated with the Homunculus and the Butterfly. 
We would like also to thank the referee 

\begin{acknowledgments}

This paper makes use of the following ALMA data:ADS/JAO.ALMA\#2016.1.00585.S. ALMA is a partnership of ESO (representing its member states), 
NSF (USA) and NINS (Japan),  together with NRC (Canada), MOST and ASIAA (Taiwan), and KASI (Republic of Korea), in cooperation with the Republic of Chile. 
The Joint ALMA Observatory is operated by ESO, AUI/NRAO and NAOJ. The National Radio Astronomy Observatory is a facility of the National Science Foundation 
operated under cooperative agreement by Associated Universities, Inc. L.A.Z. acknowledges financial support from CONACyT-280775 and UNAM-PAPIIT IN110618 
grants, México.  L.F.R., is grateful with CONACyT, México, and DGAPA, UNAM for the financial support. J.A.T. acknowledges support from UNAM-PAPIIT IA100720 
project and the Marcos Moshinsky Foundation. Part of this research is based on observations made with the NASA/ESA Hubble Space Telescope obtained from the Space 
Telescope Science Institute, which is operated by the Association of Universities for Research in Astronomy, Inc., under NASA contract NAS 5$–$26555.
We would like to thank Andrea Mehner for providing us an infrared image of $\eta$ Carinae.  We would like also to thank the referee for carefully reading our manuscript and 
for giving constructive comments, which helped improving our study.

\end{acknowledgments}

\facilities{ALMA}
\software{CASA \citep{mac2007}, KARMA \citep{goo1996}}

\bibliographystyle{aasjournal}

\appendix 

In this Appendix we include the complete velocity channel map of the CO emission toward $\eta$ Carina and a comparison of the CO (3$-$2) 
spectra taken with APEX and ALMA, which shows a small lost of emission by the interferometer.

\begin{figure*}[ht!]
\epsscale{1.0}
\plotone{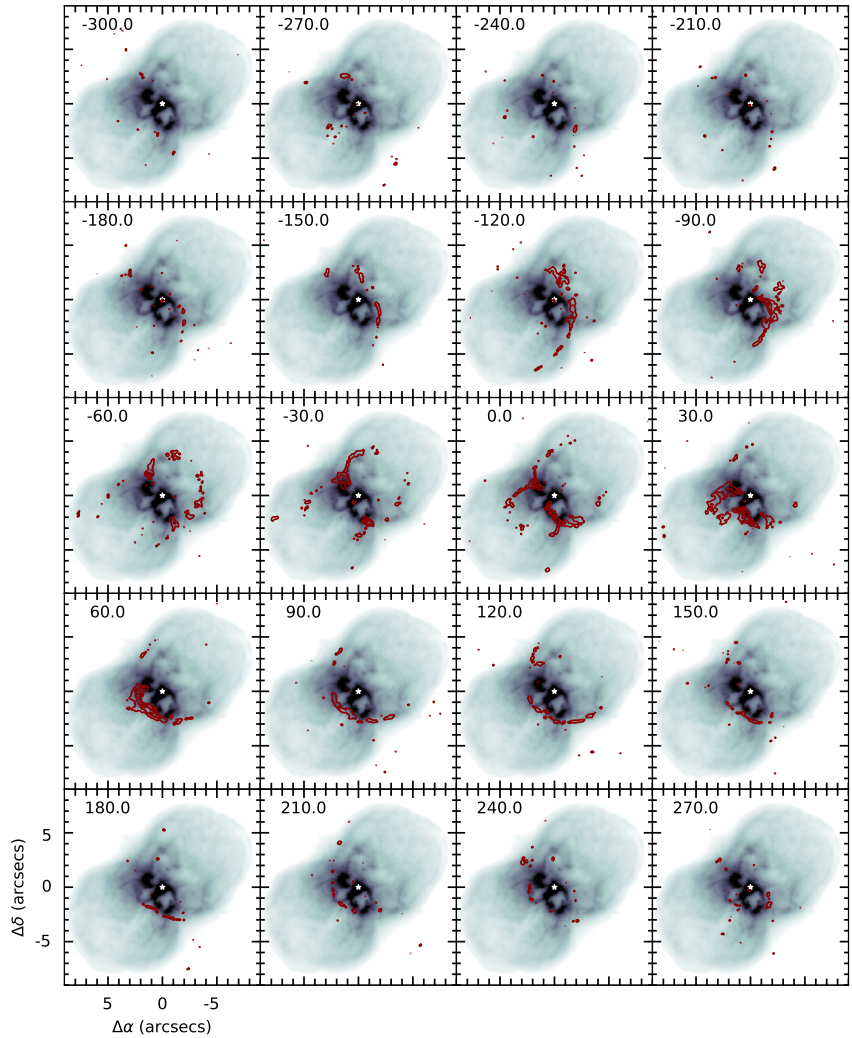}
\caption{ALMA CO(3$-$2) velocity cube (contours) overlaid infrared 7.9 $\mu$m VLT/VISIR \citep{men2019} map (grey scale) of $\eta$ Carinae. 
On top of every panel is shown the radial velocities in km s$^{-1}$. We have binned our spectral resolution to 30 km s$^{-1}$. 
The contours are $-$3, 3, 20, 40, and 60 times the sigma value in the binned cube, which is 1.8 mJy Beam$^{-1}$. The compact sources 
far from the Homunculus, and the Butterfly nebula are real and probably arising from an older and extended molecular estructure.  
\label{fig:fig5}}
\end{figure*}

\begin{figure*}[ht!]
\epsscale{1.0}
\plotone{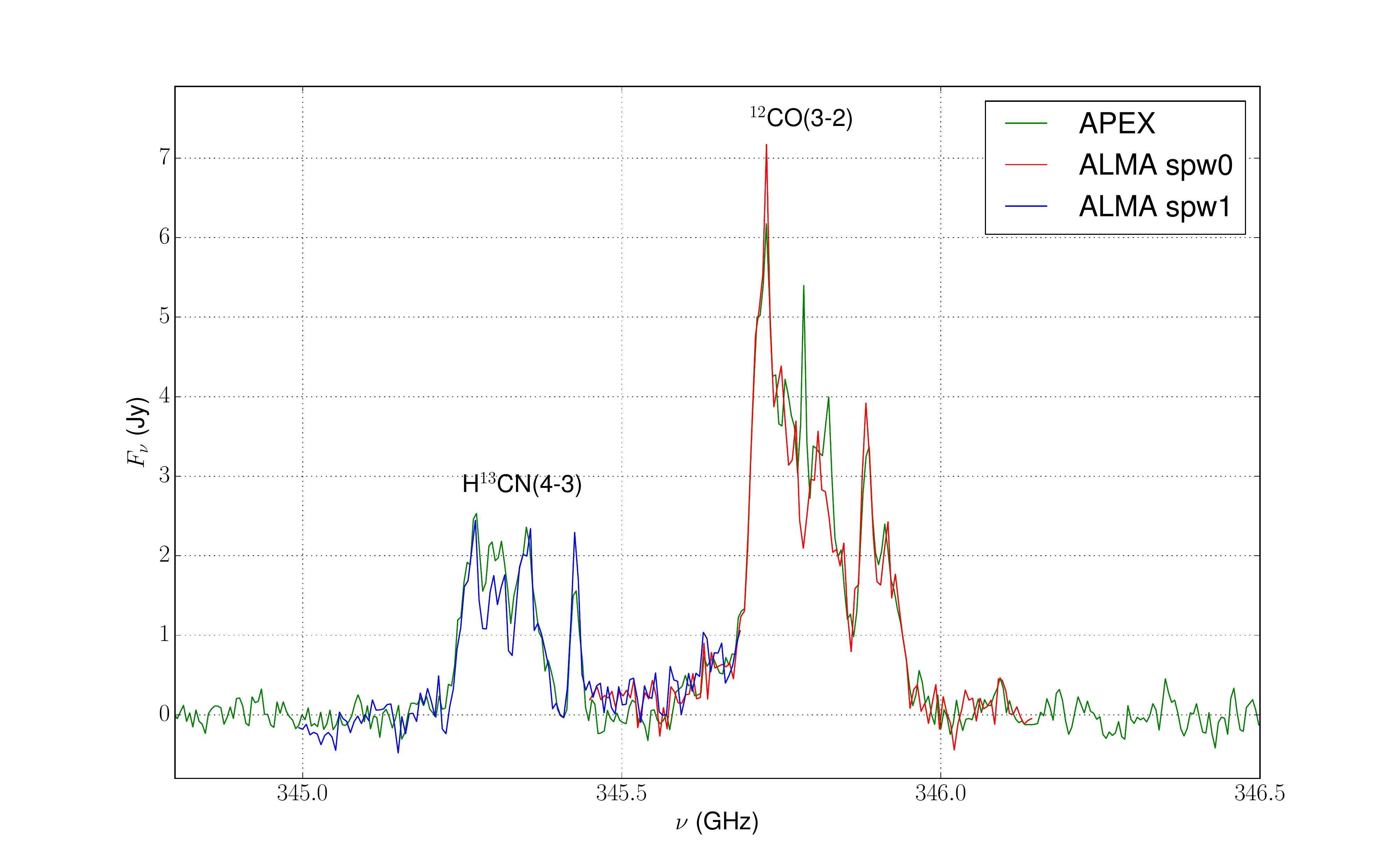}
\caption{ALMA/APEX $^{12}$CO(3$-$2) and H$^{13}$CN(4$-$3) spectra from $\eta$ Carinae.  The APEX spectra was obtained from \citet{loi2012}. 
The H$^{13}$CN(4$-$3) line emission will be discussed in a future study.
\label{fig:fig6}}
\end{figure*}

\begin{figure*}[ht!]
\epsscale{1.0}
\includegraphics[width=0.8\textwidth, angle=-90 ]{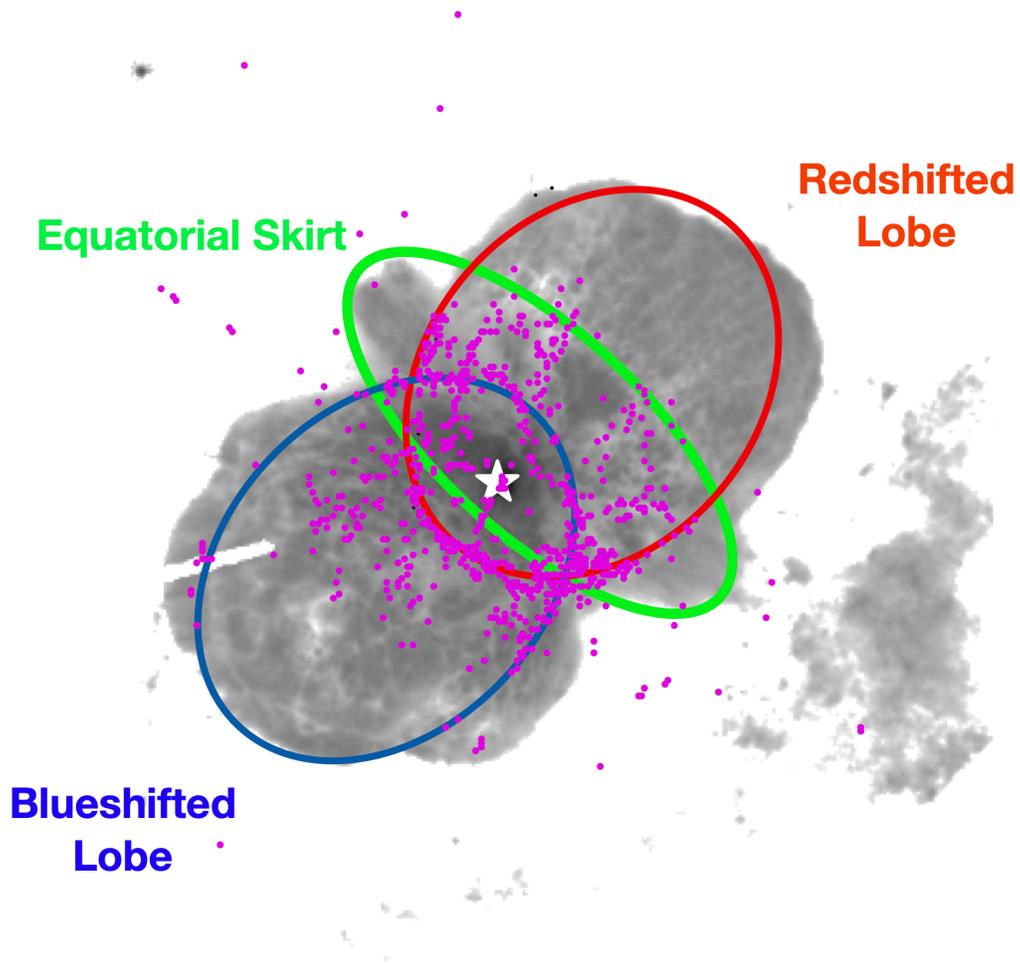}
\caption{ Archive optical HST/ACS (F660N) image (grey scale) of the $\eta$ Carinae Homunculus overlaid with the positions 
of every ALMA CO(3$-$2) condensation (violet dots) in the channel velocity cube. The white star marks the location of the binary optical star $\eta$ Carinae \citep{hog2000}.
We have fitted two ellipses (blue and red) to the high-velocity CO condensations associated with the lobes of the homunculus, and marked the position with a green ellipse 
of the optical equatorial skirt.
\label{fig:fig7}}
\end{figure*}

\begin{figure*}[ht!]
\epsscale{1.0}
\centering
\includegraphics[width=1.05\textwidth, angle=0 ]{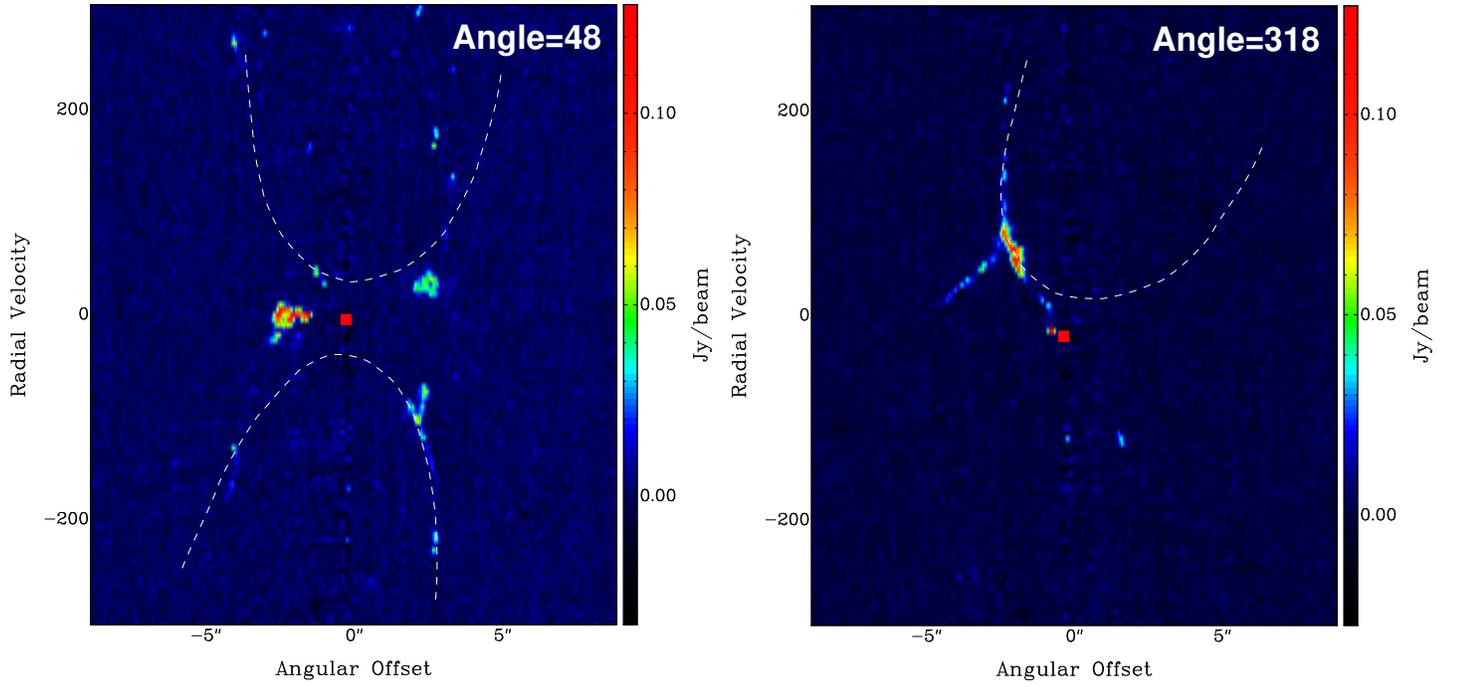}
\caption{Position-velocity diagrams of the $^{12}$CO(3$-$2) emission from the Homunculus and $\eta$ Carinae.  The position angles for the cuts are 48$^\circ$ and 318$^\circ$. 
The celestial north traces the origin of the position angles. The red square marks the location of the binary optical star $\eta$ Carinae \citep{hog2000}. The dashed line curves trace 
the blueshifted and redshifted lobes of the Homunculus Nebula. 
\label{fig:fig8}}
\end{figure*}




\end{document}